\documentclass[apjl]{emulateapj}
\usepackage{apjfonts}

\newcommand{\eg}{e.g.,\ }

\newcommand{\etal}{et~al.\ }
\newcommand{\ltsima}{$\; \buildrel < \over \sim \;$}
\newcommand{\simlt}{\lower.5ex\hbox{\ltsima}}
\newcommand{\gtsima}{$\; \buildrel > \over \sim \;$}
\newcommand{\simgt}{\lower.5ex\hbox{\gtsima}}

\newcommand{\magsec}{mag/arcsec$^2$}

\def\muv{$\mu_{\mbox{v}}$}

\begin{document}

\title{Diffuse Light in the Virgo Cluster}

\author{J. Christopher Mihos, Paul Harding, John
Feldmeier\altaffilmark{1}, and Heather Morrison}
\email{mihos@case.edu, harding@\-dropbear.\-case.\-edu, 
johnf@noao.edu, heather@\-vegemite.\-case.\-edu}
\affil{Department of Astronomy, Case Western Reserve University,
10900 Euclid Ave, Cleveland, OH 44106}

\altaffiltext{1}{NSF Astronomy and Astrophysics Fellow, NOAO Tucson}

\begin{abstract}

We present deep optical imaging of the inner $\sim$ 1.5\degr x1.5\degr\ of
the Virgo cluster to search for diffuse intracluster light (ICL). Our
image reaches a $1 \sigma$ depth of \muv=28.5 \magsec\ --- 1.5 \magsec\
deeper than previous surveys --- and reveals an intricate web of
diffuse intracluster light. We see several long ($>$100 kpc) tidal
streamers, as well as a myriad of smaller-scale tidal tails and
bridges between galaxies.  The diffuse halo of M87 is traced out to
nearly 200 kpc, appearing very irregular on these scales, while
significant diffuse light is also detected around the M84/M86
pair. Several galaxies in the core are embedded in common envelopes,
suggesting they are true physical subgroups. The complex substructure
of Virgo's diffuse ICL reflects the hierarchical nature of cluster
assembly, rather than being the product of smooth accretion around a
central galaxy.

\end{abstract}


\section{Introduction}

The diffuse intracluster light (ICL) in galaxy clusters is a valuable
tool for studying their dynamical history. First identified (see
V{\'i}lchez-G{\'o}mez 1999 for review) and quantified (\eg Uson,
Boughn, \& Kuhn 1991; Bernstein \etal 1995; Feldmeier \etal 2002,
2004) using deep broadband imaging, the ICL has also been detected
both in individual stars (Ferguson, Tanvir, \& von Hippel 1998;
Durrell \etal 2002) and intracluster planetary nebulae (IPNe; see
Feldmeier \etal 2004 and Aguerri \etal 2005 and references therein),
the latter opening up a new field studying ICL kinematics (Arnaboldi
\etal 2004; Gerhard \etal 2005). On the theoretical front, N-body
simulations of galaxy clusters are now starting to yield
high-resolution predictions of the spatial and kinematic distribution
of ICL (\eg Willman \etal 2004; Murante \etal 2004; Sommer-Larsen, Romeo,
\& Portinari 2005), and showing that
the formation of the ICL is intimately linked to tidal stripping
during the hierarchical assembly of clusters.  The phase-space
distribution of the ICL thus holds important information about the
detailed dynamical state and assembly history of a galaxy cluster.

In this context, the Virgo cluster (at an adopted distance of 16 Mpc)
is of particular interest as it shows many signs of being a
dynamically complex environment, ideal for the production of ICL. The
cluster possesses both spatial (Bingelli, Sandage, \& Tammann 1987)
and kinematic (Bingelli, Popescu, \& Tammann 1993) substructure, with
different subgroups possessing different morphological mixes of
galaxies (Bingelli \etal 1987). Many of the cluster galaxies are
kinematically disturbed (Rubin, Waterman, \& Kenney 1999), indicative
of strong, ongoing tidal interactions. Deep photographic imaging (\eg
Malin 1994; Katsiyannis \etal 1998) also show tidal features around
many Virgo galaxies, as well as an extended stellar envelope around
M87 (Weil, Bland-Hawthorn, \& Malin 1997). Studies of the ICL in Virgo
also hint at an irregular structure: the distribution of intracluster
PNe show marked field-to-field variations (Durrell \etal 2002;
Feldmeier \etal 2004; Aguerri \etal 2005), suggestive of a poorly mixed ICL.

However, quantifying the large-scale structure of the ICL in nearby
clusters like Virgo has actually proved quite difficult. The
relatively small field of view of most telescope/CCD systems makes it
hard to survey nearby clusters of large angular size.  Deep surface
photometry also is hindered by the need to accurately flat field over
degree scales, while IPNe surveys to date have not gone deep enough to
uniformly sample the full line-of-sight depth of the Virgo cluster.
Accordingly, the distribution of ICL in Virgo is still poorly
determined, yet it is here that there is a wealth of high spatial
resolution ancillary data on the properties of the galaxies and hot
intracluster medium.

To address this disparity and make {\it quantitative, large-scale
measurements} of the diffuse light in Virgo, we have begun a deep
imaging study of Virgo's core using Case's Burrell Schmidt
telescope. Because the Schmidt's wide field of view is imaged onto a
single CCD chip, highly accurate flat fielding is possible. Its
closed-tube design and extensive optical baffling significantly reduce
scattered light problems compared to open-tube telescopes. As such,
the Burrell Schmidt is optimally suited for deep, wide-field surface
photometry. Here we report on the first season of
observations which have revealed Virgo's intricate web of diffuse ICL.

\section{Imaging Technique}

\begin{figure*} \figurenum{1}
\centerline{\includegraphics[scale=0.9,angle=0]{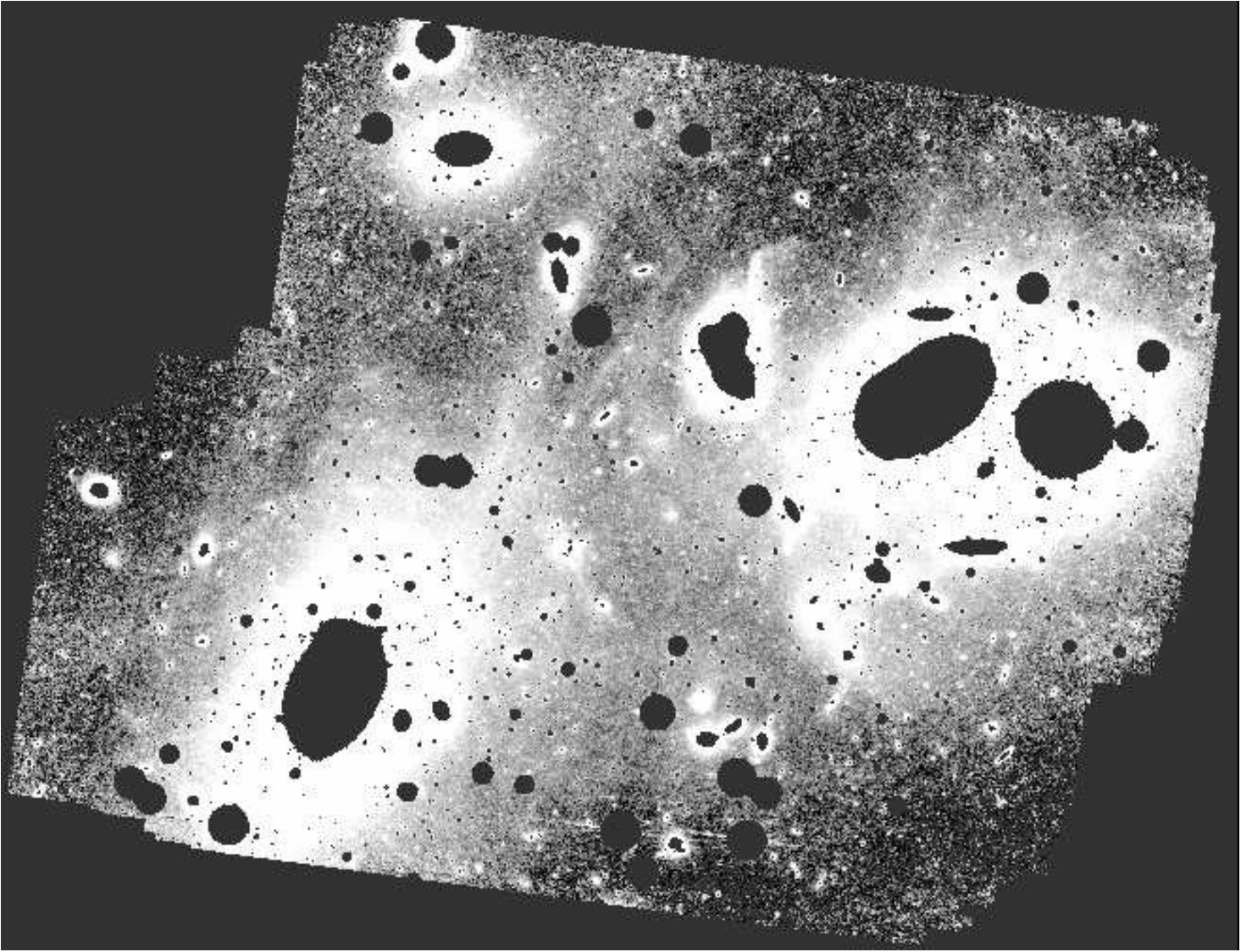}}
\caption{Diffuse light in the Virgo cluster core. North is up; east is to
the left. The white levels saturate at \muv $\sim$26.5, 
while the faintest features visible have a surface brightness of
\muv $\sim$28.5.
\label{fig1}}
\end{figure*}

Deep images of the Virgo cluster were taken during photometric nights
in March and April 2004 using the recently refurbished 0.6m Burrell
Schmidt telescope (for technical details of the refurbishment see
Harding \etal 2005 in preparation, and Feldmeier \etal 2002 for
details of the observing and data reduction strategies). The field of
view of the Schmidt and its SITe 2048x4096 CCD is approximately
1.5\degr x0.75\degr, with 1.45\arcsec\ pixels and the long axis of the
chip oriented east-west. A total of 72 object images were taken,
covering the inner 2.25 square degrees of Virgo. Individual images
were flattened using a night sky flat built from 127 offset sky
pointings bracketing the object exposures in time and position. The
object images and blank skies are all 900s exposures taken in
Washington M and transformed into Johnson V magnitudes.

Sky subtraction is a challenge when mosaicking together many images
which subtend only a portion of the cluster itself. To remove stars
and galaxies from the individual images, we first mask stars out to
where the PSF falls below 3 ADU (\muv=28), then subtract their
extended wings out to 0.5 ADU (\muv=30). IRAF's OBJMASK task is then
used to mask out all extended sources 2.5$\sigma$ above sky, which
essentially removes all bright galaxies. We then remove sky in a
multi-step process. We first rebin the masked images into smaller
images by calculating the mode in 32x32 pixel bins, and then identify
regions on the images which are located far away from any bright
galaxy and considered ``pure sky''; on average, about 5\% of each
image is covered by pure sky.  Then for each image, we calculate a DC
sky level from the mode of the counts in these regions and subtract it
from the image. This sky level ranges from 1100--1500 ADU
(\muv=21.6--21.3), depending on a variety of factors such as hour
angle, airmass, time of night, and air glow variability. Next, we perform 
an iterative plane-fitting process
where the pure sky regions are used to simultaneously fit individual
residual sky planes to each binned image, constrained to minimize
frame-to-frame deviations in the pure-sky regions of the final
sky-subtracted images. The edge-to-edge gradient of the final fitted
sky planes is typically 5 ADU, or $\sim$ 0.35\% of the DC sky
level.\footnote{It is important to recognize that fitting and
subtracting sky planes can impose a systematic underestimate of the
Virgo ICL, since any degree-scale diffuse ICL component will also
be subtracted at this step.}

After sky subtraction of the individual images, the images are
registered and medianed together to form the final mosaic. At this
point one last plane is also fit and subtracted from the mosaic using
the pure sky regions; the edge-to-edge gradient of this final plane is
2 ADU ($\sim$ 0.1\% of sky). We then improve the signal-to-noise by
medianing the mosaic in a 9x9 pixel boxes ($\sim$ 1 kpc$^2$), which
brings out the faintest features. In the raw image, the 1$\sigma$
level in the sky is 4.5 ADU (\muv=27.5), while in the binned, medianed
image 1$\sigma$=1.7 ADU (\muv=28.6). This limiting surface brightness
is approximately 1.5 \magsec\ deeper than the deep photographic
imaging of Malin (1994).

As a final note, one possible source of confusion in our study is
contamination due to the presence of any galactic cirrus in the field,
which will be visible as large-scale diffuse reflected light.
Examining the IRAS 100\micron\ maps of Virgo, it is clear that a ring
of cirrus ($\sim$ 1.5\degr\ in diameter) surrounds the Virgo core and
may contaminate fields to the southwest and north of our field.
Fortuitously, however, the Virgo core itself is quite clear of
galactic cirrus, arguing that features we detect are bona fide
structures in Virgo's diffuse light.

\section{Diffuse Light in Virgo's Core}

A greyscale representation of the final binned mosaic is shown in
Figure 1, while Figure 2 shows the image colored according to surface
brightness. The faintest features visible are at \muv $\sim$ 28.5 or
about 0.1\% of sky. A wealth of diffuse features can be seen in
Virgo's ICL, ranging from extended low surface brightness envelopes to
long, thin streamers, as well as smaller-scale tidal features
associated with many of the Virgo galaxies. Our goal here is to focus
on the qualitative morphology of the ICL; a more detailed quantitative
analysis will be addressed in later papers. Figure 3 shows a finding
chart for the features described below.

\begin{figure} \figurenum{2}
\plotone{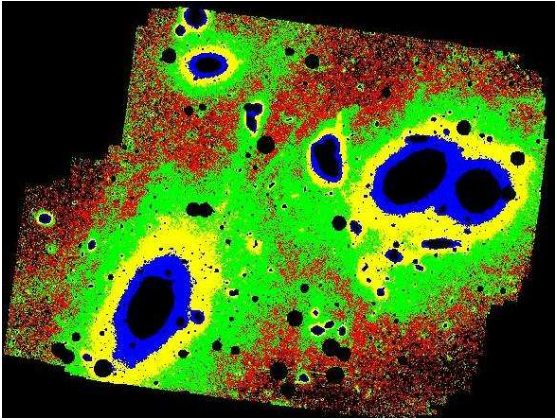}
\caption{Diffuse light in the Virgo cluster, color
coded by surface brightness: Blue: $\mu_V$=25-26;  Yellow: $\mu_V$=26-27;
Green: $\mu_V$=27-28; Red: $\mu_V$=28-29. North is up; east is to the left.
\label{fig2}}
\end{figure}

\begin{figure} \figurenum{3}
\plotone{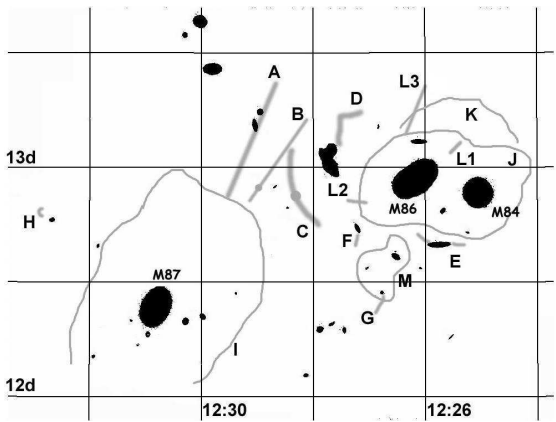}
\caption{Schematic diagram showing location of diffuse features discussed
in the text.
\label{fig3}}
\end{figure}

In the image, two long streamers can be seen extending to the
northwest from M87 at the lower left. One streamer (A) projects
through the pair of galaxies NGC 4458/61 and extends beyond, toward a
group of galaxies to the north. There is a surface brightness gradient
along its length; near M87 it has a surface brightness of \muv=27,
dropping to \muv=27.8 near the pair. From the point at which it
emerges from M87's stellar halo to where it fades into the background,
the streamer has total length of 38\arcmin\ (178 kpc) a characteristic
width of 3\farcm5 (16 kpc), and a total luminosity of $\sim 10^9\
{\rm L}_\sun$.  Because NGC 4458 and NGC 4461 have a very high
velocity relative to one another ($\Delta v = 1296$ km/s) it is
unlikely the streamer comes from any strong interaction between the
pair; instead, stripping from one of the galaxies individually is more
likely.

The second streamer (B) is perhaps even more remarkable. Also
appearing to emanate from M87 to the northwest, it measures
approximately 28\arcmin\ (130 kpc) long and only 1.2\arcmin\ (6 kpc)
wide, with a peak surface brightness of \muv=27.0. The low surface
brightness dwarf VCC 1149 is projected along the stream, just before
the position at which the stream blends into M87's halo; we may well
be seeing the destruction of this dwarf in the tidal field of the
cluster --- a process which may contribute to the variation in the
faint end slope of the luminosity function in clusters. The thinness
of the stream supports this conjecture, as the low velocity dispersion
of dwarf galaxies ensures that stripped tidal streams will be
dynamically cold, and not diffuse significantly in the tangential
direction.  The thinness and linearity of the streamer also argue that
the dwarf must be on a highly radial orbit, or one that is seen
projected along its orbital plane.

Not all detected streams are linear, however. A broad, curved stream
(C) can be seen surrounding a low surface brightness dwarf; the
curvature of this stream, arcing away from M87, suggests that the
galaxy is orbiting within or around the M86 group (or perhaps the
NGC4435/8 pair) rather than around M87 itself.

Aside from long ICL streams, many galaxies show tidal features on
smaller scales. Perhaps the most obvious is the tidal tail to the
north of the interacting pair NGC 4435/8 (D). Originally identified by
Malin (1994) in photographic imaging, our imaging traces it to fainter
surface brightnesses (\muv $>$ 28), where it takes a sudden 90\degr\
bend to the west. This kind of tidal ``dogleg'' is expected during
close and slow encounters in a galaxy cluster, where the tidal forces
of both the individual galaxies and the cluster itself conspire to
create rather complex tidal morphologies (Mihos 2005, in preparation).
Other notable features include NGC 4388's tidal plumes to the west and
northeast (E), a southern tail in NGC 4425 (F; both noted by Malin
1994) which connects to VCC 987, another tail extending from IC 3349
(G) and passing through VCC 942, and a loop to the east of IC 3475
(H).

At low surface brightness, the giant elliptical M87 shows a very
extended and irregular stellar envelope. From photographic images,
Weil \etal (1997) and Katsiyannis \etal (1998) show M87's luminous
halo extending to $\sim$ 100 kpc scales; on our image it can be traced
further out to 175 kpc (I). Here the outer isophotes are not simply
the radial extension of M87's inner regular luminosity profile; traced
to the northwest, the isophotes become boxier and show irregular loops
and shells which may be remnants from earlier stripping from, or
cannibalization of, smaller galaxies. The lopsidedness and
irregularity of M87's halo is again indicative of a diffuse envelope
which is not fully relaxed.

Because of the close proximity of M86, M84, and a number of smaller
galaxies, uniquely tracing out any individual halo in that complex of
galaxies proves difficult. The isophotes of both M86 and M84 appear
fairly regular out to a surface brightness of \muv $\sim$ 26.5 (J), as
noted by Malin (1994). At fainter surface brightnesses, the isophotes
blend together and become quite irregular, and we see a ``crown'' of
diffuse light (at \muv = 27--28) to the north of the M84/M86 system (K),
and a diffuse extension to the south of the pair, blending into
the network of galaxies to the south. Whether this extended envelope
is truly physical or simply a projection effect remains unclear.
Also visible around M86 are several low surface brightness filaments
(L1--L3), one to the northeast (\muv $\sim$ 25.5), one to the west
(\muv $\sim$ 26.5), and one to the north (\muv $\sim$ 27.5).

We also see a common envelope of light (M) surrounding the complex of
galaxies to the south of M86, including NGC 4413, IC 3363, and IC
3349, along with a tidal bridge connecting NGC 4413 and IC 3363, and
perhaps extending to NGC 4388 to the east. This envelope has fairly
irregular isophotes, arguing it is more than just simply the extended
profiles of galaxies overlapping in projection. This group of galaxies
may be a physically coherent subgroup within the Virgo cluster.

Finally, Oosterloo \& van Gorkom (2005) have recently identified an
extended (110x25 kpc) HI cloud streaming from NGC 4388 which they
attribute to ram-pressure stripping due to the lack of any stellar
counterpart in previous photographic imaging. We
also find no extended counterpart on our image down to a surface
brightness limit of \muv=28.5, save for faint streamers close to the
main body of NGC 4388 coincident with the extended emission line
region noted by Yoshida \etal (2002).

\section{Discussion}

Our deep imaging of the Virgo cluster has revealed a complex network
of extended tidal features, indicative of the ongoing stripping and
tidal evolution of galaxies in Virgo. The ICL is not radially
symmetric around M87, the nominal central galaxy of Virgo; much of the
diffuse light is centered upon the M84/M86 complex. The impact of
hierarchical assembly of clusters on the ICL is clear here --- rather
than ICL growing simply via smooth accretion around a central galaxy,
it distribution reflects the substructure inherent in the cluster, as
evidenced in cosmologically-motivated N-body simulations of ICL
(\eg Dubinski 1998; Willman \etal 2004,
Rudick \etal in preparation). Characterizing the ICL in terms of a
simple radial profile will prove misleading in all but the most
regular, relaxed clusters.

It is also instructive to compare our map of the ICL in Virgo's core
to estimates of the luminosity density derived from studies of
planetary nebulae by Feldmeier \etal (2004) and Aguerri \etal
(2005). Because IPNe detection depends strongly on the underlying
surface brightness, IPNe studies have complicated selection and
completeness functions, particularly in fields which cover both bright
galaxies and diffuse intracluster space. With this in mind, we will
restrict the discussion here to a qualitative comparison, and leave a
more quantitative analysis to a future paper.

Both IPNe studies show that the IPNe density is highly variable in the
Virgo core; our image confirms this complexity and variation in the
underlying ICL. The highest IPNe densities are found in Feldmeier
\etal\llap's ``Field 3,'' which our image shows is completely embedded
in M87's diffuse halo. Much of the IPNe detected in this field are
likely to be associated with M87 itself; radial velocity measurements
of the IPNe in this field agree with this assessment (Arnaboldi \etal
2005). Aguerri \etal\llap's ``SUB'' field also shows a relatively high
PNe density; aside from containing the luminous ellipticals M84 and
M86, this field also partly covers the diffuse envelope surrounding
the group of galaxies to the south of M86.  The lowest IPNe densities
are reported in Aguerri \etal\llap's ``LPC'' field, which also
corresponds to the lowest luminosity density in our image; indeed,
over most of this field, we detect no light above the background.

The qualitative match between the luminosity density traced by IPNe
and broadband light argues that IPne are effective tracers of the ICL
in galaxy clusters, and vice-versa. Our deep imaging provides a
``finding chart'' for IPNe searches; followup spectroscopy can then be
used to study the kinematics of discrete structures such as tidal
tails and diffuse envelopes around galaxies.  Our survey of diffuse
light in Virgo continues: we have follow-up imaging of additional
portions of the Virgo cluster, and subsequent papers will focus on
quantitative descriptions of Virgo's ICL, as well as a survey of
extremely low surface brightness galaxies in the cluster core.

\acknowledgments

This work is supported by NSF grants AST-9876143 (JCM), AST-0098435 (HLM), 
and AST-0302030 (JF), and by Research Corporation Cottrell Scholarships 
to JCM and HLM.

\end{document}